\documentclass[12pt]{article}
\usepackage{amsmath}
\usepackage{caption,subcaption,graphicx,psfrag,epsf,cases,empheq}
\usepackage{enumerate}
\usepackage{blkarray}
\usepackage{stackrel}
\usepackage{url}
\usepackage[top=1in, bottom=1in, left=1in, right=1in]{geometry} 

\usepackage{caption,subcaption,graphicx,psfrag,epsf,cases,empheq,color,amssymb,amsmath,graphicx,float,booktabs,amsthm,bm,bbm,multirow,threeparttable,textcomp,adjustbox,setspace,natbib,amsthm,soul,cancel,mathrsfs,euscript,amsfonts,xcolor,tikz,bm,blkarray,stackrel}
\usepackage{enumitem}
\usepackage{threeparttable}
\usepackage{times}
\usepackage{bm}
\usepackage{natbib}
\usepackage{siunitx}

\usepackage[plain,noend]{algorithm2e}
\usepackage{bm}
\usepackage{float}
\usepackage{tikz}
\usepackage{colortbl}
\usepackage{booktabs}
\usepackage{pbox}

\usepackage{multirow}
\usepackage{amsfonts}
\usepackage{xr}
\usepackage{xcolor, soul}
\usepackage[hidelinks]{hyperref}
\theoremstyle{plain}

\newtheorem{theorem}{Theorem}

\newtheorem{assumption}{Assumption}

\newtheorem{scenario}{Scenario}

\newtheorem*{assumption*}{\assumptionnumber}
\providecommand{\assumptionnumber}{}

\providecommand{\assumptionnumber}{}

\providecommand{\assumptionnumber}{}

\newtheorem*{example*}{\examplenumber}
\providecommand{\examplenumber}{}

\newcommand{\indep}{\perp \!\!\! \perp}

\def\bSig\mathbf{\Sigma}
\def\myinc{}

\newcommand{\Ij}{I_j}

\newcommand{\orathetak}{\oracletheta}

\newcommand{\hatincthetak}{\hat{\theta}_{\myinc k}}
\newcommand{\hatincthetaone}{\hat{\theta}_{\myinc 1}}
\newcommand{\hatincthetatwo}{\hat{\theta}_{\myinc 2}}

\newcommand{\hatincthetaj}{\hat{\theta}_{\myinc j}}
\newcommand{\hatincthetakminusone}{\hat{\theta}_{\myinc k-1}}
\newcommand{\plus}{\scalebox{0.3}{$\boldsymbol{\pmb{+}}$}}
\newcommand{\dddagger}{
  \mathbin{\vbox{\offinterlineskip\ialign{%
    \hfil##\hfil\cr
    $\plus$\cr
    $\plus$\cr
    $\plus$\cr
}}}}

\newcommand{\incgammaall}{\hat{\gamma}_{\myinc K}}
\newcommand{\incgammaallthree}{\hat{\gamma}_{\myinc K}^{\dddagger}}
\newcommand{\incgammaalltwo}{\hat{\gamma}_{\myinc K}^\ddagger}
\newcommand{\incgammaallone}{\hat{\gamma}_{\myinc K}^\dagger}

\newcommand{\incgammajone}{\hat{\gamma}_{\myinc j}^\dagger}

\newcommand{\incbetaall}{\hat{\beta}_{\myinc K}}

\newcommand{\incbetaalltwo}{\hat{\beta}_{\myinc K}^\ddagger}
\newcommand{\incbetaallthree}{\hat{\beta}_{\myinc K}^{\dddagger}}

\newcommand{\incbetaj}{\hat{\beta}_{\myinc j}}
\newcommand{\incbetajone}{\hat{\beta}_{\myinc j}^\dagger}
\newcommand{\incbetajtwo}{\hat{\beta}_{\myinc j}^\ddagger}
\newcommand{\incbetajthree}{\hat{\beta}_{\myinc j}^{\dddagger}}
\newcommand{\oracletheta}{\hat{\theta}^{\scriptstyle{ora}}}
\newcommand{\Hnjloc}{{H}_{j}}
\newcommand{\Vnjloc}{{V}_{j}}
\newcommand{\Hone}{{H}_{1}^{}}

\newcommand{\Vone}{{V}_{1}^{}}

\newcommand{\threeR}{\textsc{3r-cola}}
\newcommand{\twoR}{\textsc{2r-cola}}
\newcommand{\twoRinf}{\textsc{2r-cola-inf}}
\newcommand{\oneR}{\textsc{1r-cola}}

\newcommand{\cola}{\textsc{cola}}
\def\CP{\textsc{CP}}
\def\Fails{\textsc{Fails}}
\def\ESE{\textsc{ESE}}
\def\ASE{\textsc{MSE}}
\def\Abias{\textsc{Abias}}
\def\PS{\textsc{ps}}
\def\ATE{\textsc{ate}}
\def\INF{\textsc{inf}}
\def\T{{ \mathrm{\scriptscriptstyle T} }}

\date{}
\begin{document}
	\def\spacingset#1{\renewcommand{\baselinestretch}%
		{#1}\small\normalsize} \spacingset{1}
	\title{\Large\bf Collaborative causal inference with a distributed data-sharing management}
	\author{Mengtong Hu, Xu Shi, and Peter X.-K. Song \\
		Department of Biostatistics, University of Michigan}
	\maketitle
	
	\bigskip
	\begin{abstract}
Data sharing barriers are paramount challenges arising from multicenter clinical trials where multiple data sources are stored in a distributed fashion at different local study sites. Merging such data sources into a common data storage for a centralized statistical analysis requires a data use agreement, which is often time-consuming. Data merging may become more burdensome when causal inference is of primary interest because propensity score modeling involves combining many confounding variables, and systematic incorporation of this additional modeling in meta-analysis has not been thoroughly investigated in the literature. We propose a new causal inference framework that avoids the merging of subject-level raw data from multiple sites but needs only the sharing of summary statistics. The proposed collaborative inference enjoys maximal protection of data privacy and minimal sensitivity to unbalanced data distributions across data sources. We show theoretically and numerically that the new distributed causal inference approach has little loss of statistical power compared to the centralized method that requires merging the entire data. We present large-sample properties and algorithms for the proposed method. We illustrate its performance by simulation experiments and a real-world data example on a multicenter clinical trial of basal insulin treatment for reducing the risk of post-transplantation diabetes among kidney-transplant patients. 
	\end{abstract}
	
	\vfill
	
	\newpage
	\spacingset{1.45} 
	
\section{Introduction}

Estimation of causal effects is the central interest in the analysis of data collected from a multicenter clinical trial~\citep{hernan2002estimating}. This statistical task can be greatly challenged when serious data sharing barriers are present across multiple participating clinical centers and hard to resolve in the short run due to various logistic constraints, such as data security and privacy requirements, and tedious institutional IRB approval procedures~\citep{carter2016,mello2013,coates2020}. Such data-sharing obstacles may be significantly magnified in some trials involving international study sites. Holdups in data procurement can delay the publication of clinical findings, and consequently hold back the delivery of new  therapeutics to patients. 

Among several solutions available in the literature, meta-analysis is of great popularity. A meta-estimation of treatment effect may be calculated by an inverse-variance weighted average of site-specific treatment effects obtained from individual data sources respectively, termed the classical meta-analysis in this paper (for example,~\citep{cochran1954combination}). In this approach or others of the same meta-analytic nature, only site-specific summary statistics, rather than the full subject-level data, are involved in pooling site-specific treatment effects. However, this classical meta-analysis has two significant limitations. First, meta-analysis often concerns a single final estimator, while causal inference typically relies on intermediate steps such as building a propensity score model, the key weighting scheme of critical importance in clinical studies to balance covariate distributions across all sites~\citep{rosenbaum1983central}.
In fact, a typical meta-analysis has limited or no control over model specification at each participating site, and the pooled estimate from each site may suffer narrow interpretability, especially in the case of causal effect,  due to lack of a well-defined common estimand as well as a uniform operation for the correction of confounding effects. Second, it may suffer from data attrition due to varying recruitment capacity across study sites leading to small sample sizes and low sample variability at some study sites, which can impair the statistical power of the analysis.

Several methods have been developed to improve the classical meta-analysis approach with different objectives other than estimating causal treatment effects.  \citet{jordan2018communication} developed a surrogate likelihood framework that communicates locally estimated gradients to update the likelihood function for estimation and inference, and the convergence rate for the estimators are at the order of the local sample size. The framework is later extended by~\citet{duan2018odal,duan2019heterogeneity} for distributed clinical datasets. A distributed empirical likelihood method designed for  unbalanced datasets is recently proposed by~\cite{Zhou_2017}. Recent work by~\cite{fed2021} considers producing global propensity scores from the locally estimated gradients in a federated learning setting.~\cite{han2021federated} introduces an adaptive federated procedure of weighing the estimators locally estimated from source sites to augment average treatment effect estimate in a target site, meanwhile allowing potential heterogeneity in covariate distributions. Most of the newly proposed meta-analytic approaches adopt a divide-and-conquer strategy, similar to a parallelized operation that requires reliable local estimates and inferential quantities. Unfortunately, due to various reasons pointed out above, the demand for high-quality numerical results from all local sites is rarely satisfied in practice.

This paper focuses on the development of a more flexible and reliable meta-analysis methodology by overcoming the above-marked impediments to evaluating causal treatment effects through effective data-sharing management.  Among multiple possible strategies concerning the calculation and utilization of propensity scores in the case of distributed data, we design four new procedures for effectual communication of summary statistics across study sites. Consequently, we can seamlessly integrate the propensity score calculation and causal treatment effect estimation using a systematic cross-site collaboration.  We term this new approach as \textit{collaborative operation of linked analysis  (\cola)}. 

\begin{figure}

	\includegraphics[width=\textwidth]{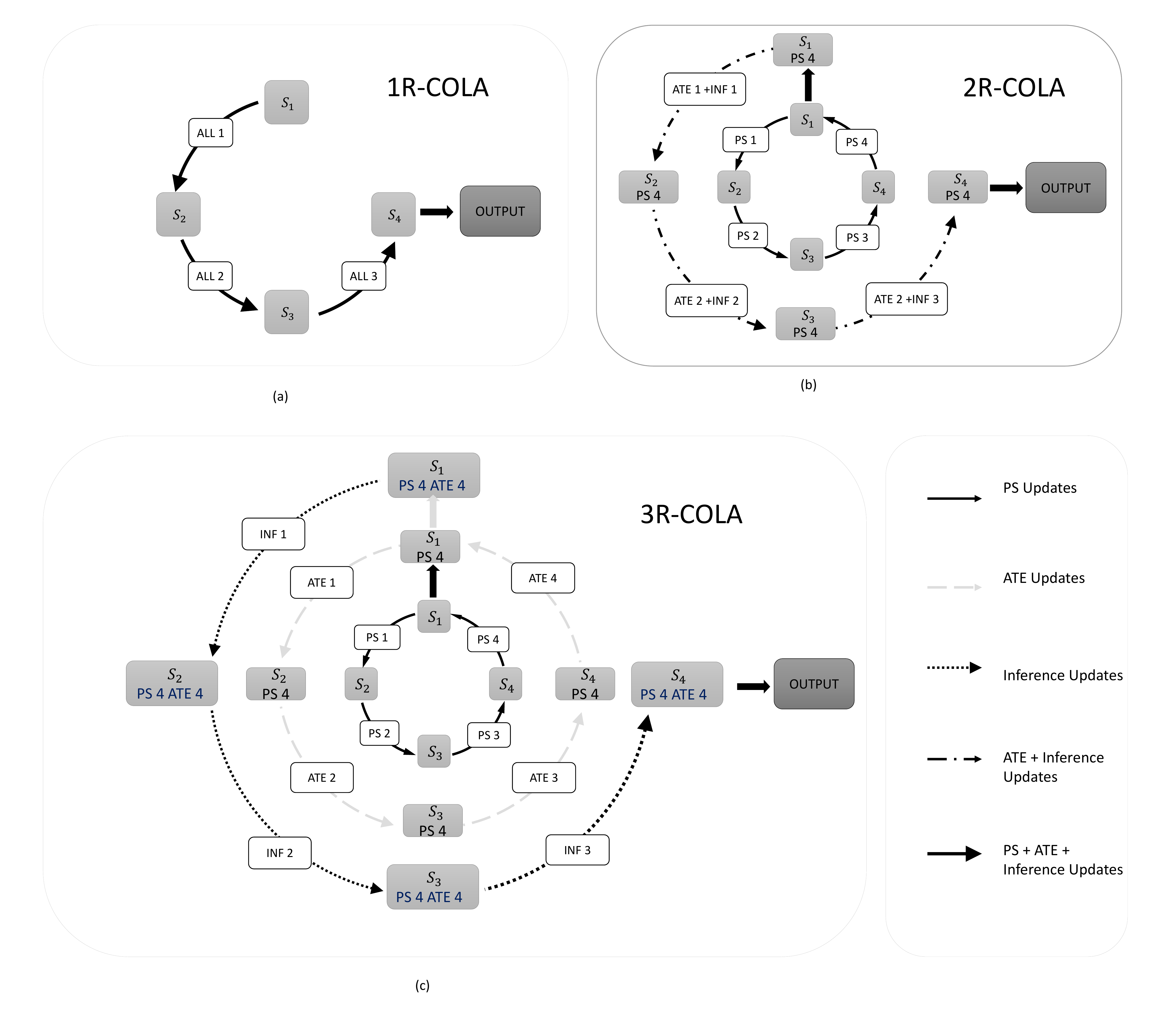}
    \caption{A diagram showing \oneR, \twoR, and \threeR\ procedures of the collaborative causal inference framework. Different types of arrows indicate the major updates involved in each round as shown in the legend. ATE which is short for Average Treatment Effect denotes the causal effect estimate.  
    Panel a) shows \oneR\ evolves one round operation that updates PS, ATE, and inferential quantities simultaneously. Panel b) shows \twoR\ involving two rounds of updates where the first round produces PS estimates, and the second round produces a causal effect and its variance. Panel c) shows \threeR\ involving three-rounds of updates.}
    \label{fig:my_label}
\end{figure}

The reason that the \cola\ method appears more flexible than existing meta-analysis is that it adopts a serial updating machinery~\citep{Luo_2020}, different from the currently popular parallelized operation, in the cross-site communication of summary statistics. The improved flexibility is achieved by different options of passing information in the \cola\ machinery to reach a desirable trade-off between information communication cost and statistical estimation efficiency. Figure \ref{fig:my_label} shows our proposed relays for summary data communications, which leads to a fully efficient inverse probability weighting estimation of treatment effect in the sense that its convergence rate is at the order of the cumulative sample size. We show both theoretically and numerically that the \cola\ has no loss of statistical power in comparison to the oracle estimation obtained by the centralized analysis that merges data from all sites.

To elucidate key elements and insights of the proposed \cola\ methodology, we choose a simple but practically important setting of logistic regression with binary outcomes, which is largely motivated by a multicenter clinical study of the Insulin Therapy for the Prevention of New-Onset Diabetes after Transplantation (ITP-NODAT) trial. This cross-border clinical study is a multi-center randomized clinical trial conducted at four kidney transplant centers in Barcelona, Berlin, Graz, and Vienna, respectively. The goal of the study is to estimate the average treatment effect of basal insulin intervention on preventing overt post-transplantation diabetes mellitus (PTDM) using data collected from the four hospitals. According to \citet{Schwaiger_2021}, the collected data exhibit strong imbalances in some covariate distributions and logistic challenges in the creation of a centralized database.  Meta-analysis is not the choice to bypass the need for data merging because zero disease cases are observed from the treatment group in Barcelona and likewise, from the control group in Graz. Thus, these two hospitals cannot provide local estimates, and the classical meta-analysis would use results only from two sites (Vienna and Berlin) and would thus be clearly underpowered.  To expedite the delivery of clinical findings, it is of great interest to extend the capacity of existing meta-analysis for the needed flexibility of embracing rare outcomes and balancing covariate distributions to retain statistical power and produce an efficient and reliable estimation of the causal effect of the insulin therapy. With little effort, the \cola\ framework developed for the binary outcome may be extended to other continuous and categorical outcomes in the context of generalized linear models.   

The organization of this paper is as follows. Section~\ref{sec:Methodology} begins with formulation and model assumptions for estimating causal effects and then introduces the proposed \cola\ framework. Section~\ref{sec:Implementation} presents different options for passing information in the \cola\ machinery. Section~\ref{sec:Asymptotics} establishes the theoretical guarantees for our proposed methods. We illustrate our proposed methods with simulation studies in Section~\ref{sec:simulation} and an application to the ITP-NODAT data example in Section~\ref{sec:data application}. We make some concluding remarks in Section~\ref{sec: discussion}.

\section{Basic setup}\label{sec:Methodology}
Consider a random sample of $N$ individuals independently sampled from $K$ clinical sites, indexed by $j = 1, \cdots , K$, each site having a sample size of $n_j$. 
For each individual $i$, we observe an outcome  $Y_{i}$ of interest
, a binary treatment indicator $A_{i} \in \{0, 1\}$, and a vector of baseline covariates $X_{i}$. Let $\Ij = \{n_{j-1}+1,\ldots, n_{j-1}+n_j\}$  to be the index set for subjects from the $j$th site 
and we denote the $j$th site-specific data as $S_j=\{(Y_{i}, A_{i},X_{i}): i\in \Ij\}$. We assume 
 $\{(Y_{i}, A_{i},X_{i}): i\in \bigcup_{j=1}^K \Ij \}$ are independent and identical observations under the same underlying population. Following the potential outcome framework \citep{neyman1923application,rubin2005}, we assume that there exists a pair of potential outcomes $\{Y_{i}(1), Y_{i}(0)\}$ for each individual $i$, representing the outcomes had the individual received treatment $(1)$ or control $(0)$.  The causal effect is typically defined as a contrast between $\mu_1=E\{Y(1)\}$ and $\mu_0=E\{Y(0)\}$, such as the mean difference, $\Delta_{D} = \mu_1 - \mu_0$, the causal log risk ratio (logRR),  $\Delta_{RR}= log \left({\mu_1}/{\mu_0}\right)$, and the causal log odds ratio (logOR), $\Delta_{OR}= log[\{{\mu_1}/{(1-\mu_1)}\}/\{{\mu_0}/{(1-\mu_0)}\}]$. To proceed,  we postulate the following four standard assumptions for the identification of the causal estimands:
\begin{assumption}[Causal effect identification]\label{a:causal} For $a\in\{0,1\}$, 
\begin{enumerate}[label = (\alph*), ref = \ref{a:causal}(\alph*)]
\item \label{a1:con} Consistency: $Y=Y(a)$ almost surely
when $A = a$ .
\item\label{a1:no} Ignorability: $\{Y(0), Y(1)\} \indep  A\mid X$.
\item\label{a1:pos}Positivity:
$0<\mbox{pr}(A=a\mid X)<1$ for all $a$ almost surely .

\end{enumerate}

\end{assumption}

 Under Assumption~\ref{a:causal}, the mean potential outcome is identified as $E[Y(a)]=E[E(Y\mid A=a,X)]=E[I(A=a)Y/P(A=a\mid X)]$, and can be estimated through different methods including the inverse probability of treatment weighting (\textsc{IPTW}), G-computation, and augmented inverse propensity weighting (\textsc{AIPTW})~\citep{robins1994estimation}.

\subsection{Inverse Propensity Treatment Weighting estimator }\label{subsec:2.1}
In this paper, we illustrate our method using IPTW estimation, although our method also applies to other causal estimation methods such as G-computation and AIPTW. The propensity score $e(X)$ is the probability of being assigned to the treatment group conditional on the covariates, i.e.,  $e(X)=\mbox{pr}(A=1\mid X)$. The propensity score model for $e(X)$ used in IPTW method needs to be correctly specified for consistent estimation of the causal effect. As such we introduce the following assumption.  
\begin{assumption}[Propensity score model]
\label{a:model}
 A propensity score model $e(X;\gamma)$ with a finite-dimensional parameter $\gamma$  is correctly specified for $e(X)$. 
\end{assumption}
Typically, the propensity score is modelled and estimated by the logistic regression, namely $e(X;\gamma)=\text{logit}(X^\T\gamma)$. We begin with a brief review of the classical estimation and inference method based on \textsc{IPTW}
in the setting where data from all sites are available and analyzed in a centralized fashion. We term this situation as the \emph{oracle} setting in this paper.
The IPTW method is a two-stage procedure. First, we estimate $\gamma$ as the solution to the logistic model equation  $\sum_{i=1}^{N} \Psi^{\text{\text{ps}}}(A_i,X_i;\gamma)=0 $, denoted by $\hat{\gamma}$, where the kernel function is, suppressing index $i$, 
$\Psi^{\text{ps}}(A,X;\gamma) = X\{A-e(X;\gamma)\}.
$ Then, we fit a marginal structural model by solving  $\sum_{i=1}^{N}  \Psi^{\text{$\Delta$}}(A_i,X_i,Y_i;\hat{\gamma},\beta)=0$, with 
$
    \Psi^{\text{$\Delta$}}(A,X,Y;\gamma,\beta) = (1, A)^\T \omega(A,X;\gamma) \{Y- g(\beta_0+\beta_{\scriptscriptstyle A} A)\}
$ 
, where the weight is  $\omega(A,X;\gamma)=A/e(X;\gamma)+(1-A)/\{1-e(X;\gamma)\}$. The link function  $g(.)$ is a known link function~\citep{mccullagh2019generalized}. For example, when the outcome is  binary, one typically uses the logistic link function $g(x) = 1/(1+e^{-x})$ and the slope parameter, i.e the coefficient for treatment $\beta_{\scriptscriptstyle A}=\Delta_{OR}$. 
To establish a valid inference, we jointly estimate all model parameters related to propensity scores and the causal effects.
Let $\theta=(\gamma,\beta)^\T$ where true values are $\theta_0=(\gamma_0,\beta_0)^\T$. Stacking $\Psi^{\text{ps}}$ and $\Psi^{\text{$\Delta$}}$, we form a joint estimating function
 \begin{equation}
 \label{one}
\Psi(\theta)=\Psi(A,X,Y;\theta)=\begin{pmatrix}
\Psi^{\text{ps}}(A,X;\gamma)\\ \Psi^{\text{$\Delta$}}(A,X,Y;\gamma,\beta)
\end{pmatrix}.
 \end{equation}
Under Assumption~\ref{a:causal} and~\ref{a:model}, the estimating function has mean zero when evaluated at the true value $\theta_0$, i,e.,  $E\{\Psi(\theta_0)\}=0$ which will be needed for deriving the asymptotic results in Section~\ref{sec:Asymptotics}.

Let $\oracletheta$ denote the oracle estimator obtained from the centralized analysis which solves \\
 $\sum_{i=1}^N \Psi_i(\theta)=0$, where $\Psi_i(\theta)=\Psi(A_i,X_i,Y_i;\theta)$. 
 Since $ \Psi(\theta)$ is an unbiased estimating function, under some mild regularity conditions, we have the asymptotic normality for $\orathetak$, namely
$\surd{N}(\orathetak-\theta_0)\xrightarrow{d}N(0,J(\theta_0))$
where $J(\theta_0)=H(\theta_0)^{-1} V(\theta_0)\left\{H(\theta_0)^{-1}\right\}^\T$, where the variability matrix $V(\theta_0)=E_{\theta_0}\{\Psi(\theta_0)\Psi^\T (\theta_0)\}$, and the sensitivity matrix $H(\theta_0)=-E_{\theta_0}\{{\partial \Psi(\theta_0)/ \partial \theta^\T}\}$.  $J$ is the inverse of the Godambe information matrix also named as the sandwich covariance matrix 
~\citep{godambe1960,godambe1991,Stefanski_Boos_2002,song2007}.
We can estimate this asymptotic covariance  using their sample counterparts given by 
${V}({\orathetak})=\sum_{i=1}^{N} {\Psi_i(\theta)\Psi_i^\T (\theta)\bigl\vert_{\theta=\oracletheta}} $, and  ${H}({\orathetak})=\sum_{i=1}^{N}  {-\partial \Psi_i(\theta)/{\partial \theta^\T}\bigl\vert_{\theta=\orathetak}}.$
~The resulting centralized oracle estimator $\oracletheta$ and its covariance serve as the gold standard to contrast the performance of our proposed distributed methods. 
\subsection{Incremental causal effect estimator}

\label{subsec:incremental}
We consider a situation of practical importance where pooling data from multiple sites is prohibited. To address this data-sharing challenge, we propose a regularized estimation method, termed Collaborative Operation of Linked Analysis (\cola), which does not require sharing individual-level data but only certain summary statistics across institutes. Note that \cola\ is not derived in a parallel computing paradigm, different from most of the existing solutions. Specifically, given an order of study sites, an incremental estimator obtained at site $k$, denoted by $\hatincthetak$, is sequentially updated over the first $k$ sites, beginning with $\hatincthetaone$ at study site 1. Obviously,  $\hatincthetaone$ is the same as a local estimator as a root of the estimating equation $\sum_{i\in I_1} \Psi_i(\theta)=0$ with the local data of $S_1$.
We define $$\Hnjloc(\hatincthetaj)=\sum_{i\in I_j} {-\frac{\partial \Psi_i(\theta)}{\partial \theta^\T}\bigl\vert_{\theta=\hatincthetaj}},~ \Vnjloc(\hatincthetaj)=\sum_{i\in I_j} {\Psi_i(\theta)\Psi_i^\T (\theta)\bigl\vert_{\theta=\hatincthetaj}}$$ as the sensitivity and variability matrix, respectively, evaluated at a local site $j\in\{1,\dots,K\}$.
The initial estimates of the sensitivity matrix and variability matrix 
$\{\Hone(\hatincthetaone)$, $\Vone(\hatincthetaone)\}$ with the local data $S_1$, are also updated by \cola. 
After obtaining $\{\hatincthetaone$, $\Hone(\hatincthetaone)$, $\Vone(\hatincthetaone)\}$ from site 1,  \cola\ passes these summary statistics to site $2$ where the triplet updates $\hatincthetaone$ to $\hatincthetatwo$ by solving the following estimating equation~\citep{Luo_2020}:
\begin{equation*}
\Psi_{n_2}(\hatincthetatwo)+
\Hone(\hatincthetaone)(\hatincthetaone-\hatincthetatwo)
=0,
\end{equation*}
where $\Psi_{n_j}(\hatincthetaj)=\sum_{i\in I_j} \Psi_i(\hatincthetaj)$. Repeating this sequential updating, \cola\ can be carried out over a sequence of all sites to produce estimators and inferential quantities. 
In particular, when updating $\theta^{k-1}$ to $\theta^k$ at site $k$, we solve for a root of the following estimating equation:  
\begin{equation}
\label{two}
\Psi_{n_k}(\hatincthetak) +\sum_{j=1}^{k-1}\Hnjloc(\hatincthetaj) (\hatincthetakminusone-\hatincthetak)=0.
\end{equation}
 The estimating function in \eqref{two} 
consists of two parts: the first term $\Psi_{n_k}(\hatincthetak)$ is based on the local data $S_k$ at site $k$, and the second term assembles the cumulative summary statistics preceding from all previous $k-1$ sites. The Newton-Raphson algorithm is applied to numerically find a solution $\hatincthetak$. 

For statistical inference, we sequentially compute the cumulative sensitivity and variability matrices over the first $k$ sites evaluated at a given point estimate. For example, if we update the sensitivity and variability matrices along with the incremental estimates, then at site $k$, the sensitivity matrix is given by $ \sum_{j=1}^k\{\Hnjloc(\hatincthetaj)\} $ and the variability matrix is given by $ \sum_{j=1}^k\{\Vnjloc(\hatincthetaj)\}.$ Then we compute the estimated covariance matrix using the sandwich formula.

\section{Implementation}\label{sec:Implementation}
\begin{figure}[b]
\centering
\begin{tikzpicture}[x=0.75pt,y=0.75pt,yscale=-1,xscale=1]
\usetikzlibrary{shapes.geometric, arrows}

\tikzstyle{startstop} = [rectangle, rounded corners, minimum width=2cm, minimum height=1cm,text centered, draw=black, fill=gray!5]

\tikzstyle{startstopwide} = [rectangle, rounded corners, minimum width=3.5cm, minimum height=1cm,text centered, draw=black, fill=gray!5]
\tikzstyle{startstopwidewide} = [rectangle, rounded corners, minimum width=4cm, minimum height=1cm,text centered, draw=black, fill=gray!5]
\tikzstyle{io} = [trapezium, trapezium left angle=70, trapezium right angle=110, minimum width=3cm, minimum height=1cm, text centered, draw=black, fill=blue!30]
\tikzstyle{n} = [rectangle, rounded corners, minimum width=2cm, minimum height=1cm,text centered, draw=black]
\tikzstyle{decision} = [diamond, minimum width=3cm, minimum height=0.5cm, text centered, draw=black, fill=green!30]
\tikzstyle{arrow} = [thick,->,>=stealth]

\node(three) [n]{$\threeR$};
\node (threeR1) [startstop,right of = three, xshift = 1.5cm,label=above: \small R1: \PS] {$\incgammaallthree$};
\node (threeR2) [startstop,right of = threeR1, xshift = 1.5cm,label=above: \small R2: \ATE] {$\incbetaallthree$};
\node (threeR3) [startstop,right of = threeR2, xshift = 3.4cm,label=above:\small  R3: \INF] {
 $\sum_{j=1}^K\Hnjloc(\incgammaallthree,\incbetaallthree) $, $ \sum_{j=1}^K\Vnjloc(\incgammaallthree,\incbetaallthree)$};

\node(two) [n,below of = three,  node distance = 1.7cm]{$\twoR$};
\node(twoR1)[startstop,right of = two, xshift = 1.5cm,label=above: \small R1: \PS]
{$\incgammaalltwo$};
\node(twoR2)[startstop,right of = twoR1, xshift = 4.8cm,align =center,label=above:\small  R2: \ATE + \INF]{  $\incbetaalltwo, \quad  $ $\sum_{j=1}^K\Hnjloc(\incgammaalltwo, \incbetajtwo) $, $\quad \sum_{j=1}^K\Vnjloc(\incgammaall, \incbetajtwo)$};
\node(one) [n,below of = two,  node distance = 1.7cm]{ $\oneR$};
\node(onefull)[startstopwidewide,right of = one, xshift = 6cm,label=above: \small R1: \PS +\ATE +\INF ,minimum width=11cm]
{$\incgammaallone$, $\incbetaall^\dagger$,$\sum_{j=1}^K\Hnjloc(\incgammajone, \incbetaj(\incgammajone)) \mbox{, and }$ $\sum_{j=1}^K\Vnjloc(\incgammajone, \incbetaj(\incgammajone))$};
\node(variables)[startstopwide,below of = onefull,  node distance = 3cm,align =left,xshift = -1cm]{$\incgammaallone$ = $\incgammaalltwo$ = $\incgammaallthree$ : fully updated propensity score estimates at the last site.\\$\incgammajone$ : incrementally updated propensity score estimates at the $j$th site.\\ $\incbetaalltwo=\incbetaallthree =\incbetaall(\incgammaallthree)$ : fully updated causal treatment effect at the last site.  \\ $\incbetajtwo = \incbetaj(\incgammaalltwo)$ : partially updated causal treatment effects using fully updated prop-\\\quad \quad\quad\quad ensity score estimates at the $j$th site.\\ $\incbetaall^\dagger$ : partially updated treatment effect estimate at the last site with $\incgammajone$ plug- \\\quad \quad ged in at each preceding site. \\ $\incbetaj(\incgammajone)$ : incrementally updated treatment effect estimate at the $j$th site using $\incgammajone$. 
};
\end{tikzpicture}
\caption{Final outputs obtained after each round of update by four \cola\ methods} \label{method_compare}
\end{figure}
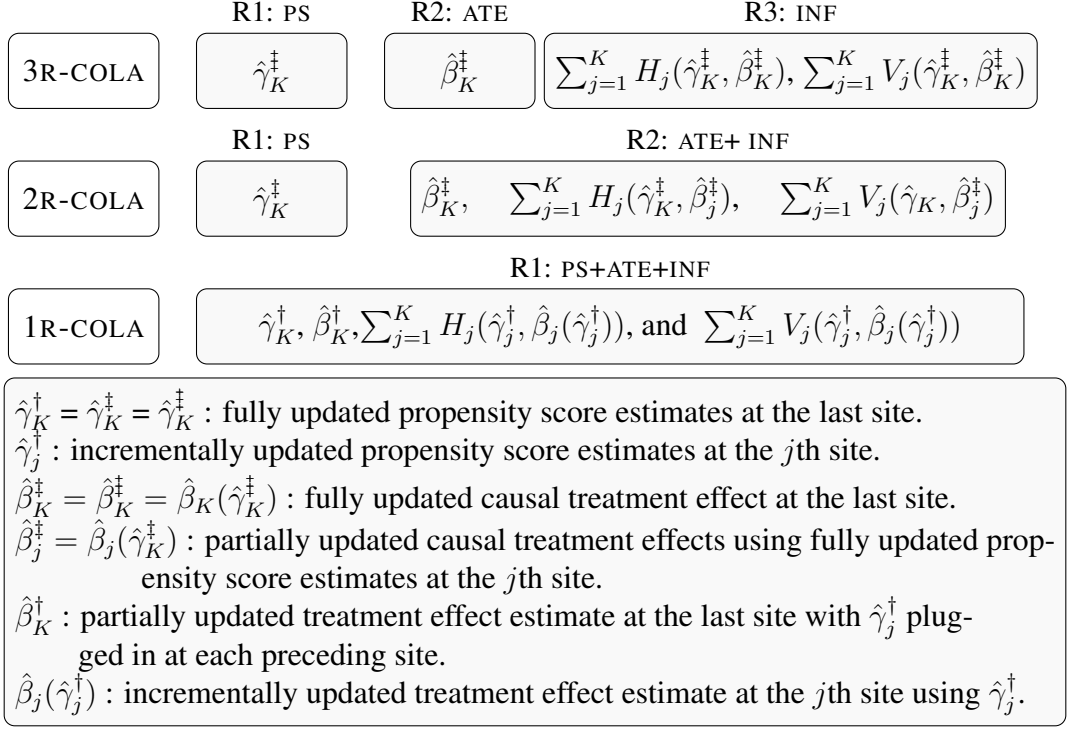
This section illustrates three ways to implement the incremental causal estimator as defined in Section \ref{subsec:incremental}, where the main nuance lies in the trade-off between communication efficiency and finite-sample numerical accuracy. The outputs at each round of updates for the three procedures introduced below are summarized in Fig \ref{method_compare}. For ease of illustration, we vary the numbers of ``+" in the subscripts of the  estimator to correspond to the number of rounds used in each implementation of \cola.

\subsection{A three-round algorithm}
An accurate estimate $\hat{\gamma}$ in the propensity score model is essential for accurate estimation and inference of the causal effect which is related to the performance of logistic regression in propensity score estimation. To implement \cola, we propose a three-round estimation algorithm, denoted by \threeR, for the population-level causal effect estimation, as shown in Fig. \ref{fig:my_label}(c).
\begin{enumerate}
\item[]{Round 1:} The first round fits the propensity score model using our sequential method and produces a ``global" estimate of the model parameter, executing a full round of sequential updating through all $K$ sites. We output the coefficient estimate $\incgammaallthree$ at the last site $K$. By Theorem 3 in Section \ref{sec:Asymptotics}, this  $\incgammaallthree$ approximates the oracle estimate at the order of $o_p({N}^{-1/2})$.
\item[]{Round 2:} The second round estimates the causal effect by the same method. We communicate $\incgammaallthree$ back to all local sites so that equation \eqref{two} can sequentially update the causal effect $\incbetajthree = \incbetaj(\incgammaallthree)$. This round outputs the global estimate of causal effect,  $\incbetaallthree =\incbetaall(\incgammaallthree),$ which is communicated back to all sites. 
\item[]{Round 3:} The third round estimates the asymptotic covariance by updating the cumulative sums $\sum_{j=1}^K\Hnjloc(\incgammaallthree,\incbetaallthree) $ and  $ \sum_{j=1}^K\Vnjloc(\incgammaallthree,\incbetaallthree)$ sequentially over all $K$ sites. 
\end{enumerate}
\vspace{-0.5pc}
\subsection{A two-round algorithm}
The \threeR\ algorithm above may be simplified to a two-round algorithm, denoted by \twoR\ illustrated in Fig. \ref{fig:my_label}(b). It combines ``Round 2" and ``Round 3" of \threeR\ for operation, with some details below. The pseudo-code for \twoR\ is detailed in the supplementary material.
\begin{enumerate}
\item[]{Round 1:} The same ``Round 1" of \threeR\ is used to output $\incgammaalltwo$ which is the same as $\incgammaallthree$.
\item[]{Round 2:} While updating $\incbetajtwo =\incbetaj(\incgammaalltwo) $, the covariance is updated simultaneously using current $\incbetajtwo$, namely $\sum_{j=1}^k\Hnjloc(\incgammaalltwo, \incbetajtwo) \mbox{ and } \sum_{j=1}^k\Vnjloc(\incgammaalltwo, \incbetajtwo)$. Obviously, the current update $\incbetajtwo$ differs from the output from \threeR\ $\incbetaallthree$. The numerical performance of the resulting inference would be different and is illustrated in simulation results.
\end{enumerate}
\twoR\ and \threeR\ produce the same point estimates for propensity scores model coefficients and causal effect, but the different covariance estimations.
Using a fully updated $\incbetaallthree$ in the covariance calculation may gain some numerical stability than that of the concurrent estimator $\incbetajtwo$ using preceding data information available in the incremental updating paradigm. 
An alternative two-round algorithm named \twoRinf\ estimates $\beta$ and $\gamma$ simultaneously in ``Round 1" and only computes and updates covariances in a ``Round 2". The details for \twoRinf\  are given in the supplementary material.

\subsection{A one-round algorithm}
 Considering minimizing communication cost, we may further simplify the updating procedure by combining all three rounds into a one-round \oneR\ algorithm shown in Fig.~\ref{fig:my_label}(a). It minimizes between-site communication, at the price of reduced numerical stability and finite-sample performance. Instead of communicating global $\incgammaallone$ back into equation~\ref{two}, concurrent estimate $\incgammajone$ is used to compute $\incbetajone(\incgammajone)$ within the same round. Concurrently, the covariance is updated through  $\sum_{j=1}^K\Hnjloc(\incgammajone, \incbetajone(\incgammajone)) \mbox{, and }$ $\sum_{j=1}^K\Vnjloc(\incgammajone, \incbetajone(\incgammajone))$.  At the last site,  \oneR\ produces point estimate $\incgammaallone \mbox{ and } \incbetaall^\dagger=\incbetaall(\incgammaallone)$, with a different covariance estimate compared to \twoR\ and \threeR.

\section{Large-sample Properties}\label{sec:Asymptotics}
Let $N_k$ be the cumulative sample size for the first $k$ sites, $N_k = \sum_{j=1}^k n_j$. We choose \oneR\ to discuss the large sample proprieties of our incremental estimators as $N_k=\sum_{j=1}^k n_k \to \infty$ instead of $\emph{min}_{j \in \{1,\cdots ,k\}}n_j \to \infty$ in the parallel computing paradigm. This condition is satisfied when $n_k \to \infty$ at one of the sites, or when the number of sites $k \to \infty$ and the former is the focus of this paper. 
The asymptotic properties of $\cola $ methods with more than one round can be minimally established with analytic effort. Thus, their proofs are omitted. Denote the $L^2$-norm of a vector $u$ by $\Vert u \Vert$.  
Let $N_{\rho}(\theta_0)=\{\theta:\|\theta-\theta_0\| \leq \rho\}$, $\rho > 0$ be a compact neighborhood of size $\rho$ around the true value $\theta_0$. In addition to Assumptions~\ref{a:causal} $-$~\ref{a:model} for causal effects identifiability and estimatability, we assume the following regularity conditions for the estimating function $\Psi$ given in equation \eqref{one} to establish some key asymptotic properties.

\begin{assumption}[Regularity Conditions]\label{a:reg} We assume the following on estimating function in equation~\ref{one}
\begin{enumerate}[label = (\alph*), ref = \ref{a:reg}(\alph*)]
\item \label{c:one}
The true value $\theta_0$ is the unique solution to $\lambda(\theta)=E\{\Psi(\theta)\}=0$.
\item \label{c:two}
The estimating function $\Psi(\theta)$ is continuously differentiable for all $\theta$ in the neighborhood $N_{\rho}(\theta_0)$.
\item \label{c:three}
The sensitivity matrix $H(\theta)$ and the variability matrix $V(\theta)$ are positive definite for all $\theta \in N_{\rho}(\theta_0)$.
\item \label{c:four}
The sensitivity matrix $H(\theta)$ is Lipschitz continuous for all $\theta \in$  $N_{\rho}(\theta_0)$.
\end{enumerate}

\end{assumption}

Conditions~\ref{c:one} $-$~\ref{c:four} are mild regularity conditions needed for legitimate asymptotic behaviors of the \cola\ estimator $\hatincthetak$ under the classical theory of estimating functions~\citep{song2007,tsiatis2006}. Condition~\ref{c:four} is satisfied usually for the generalized linear models~\citep{mccullagh2019generalized}.

\begin{theorem}
\label{theorem1}
Under the regularity conditions~\ref{c:one} $-$~\ref{c:four}, the \cola\ estimator $\hatincthetak$ is consistent for the true value $\theta_0$, i.e. $\hatincthetak\xrightarrow[]{p}\theta_0$, as $N_k \to \infty$.
\end{theorem}
\begin{theorem}
\label{theorem2}
Under the regularity conditions~\ref{c:one} $-$~\ref{c:four}, the \cola\ estimator  $\hatincthetak$ is asymptotically normally distributed, i.e., 
$\surd {N_k}(\hatincthetak-\theta_0) \xrightarrow[]{d} N(0,J(\theta_0)), \,\,\, \text{as }  N_k \to \infty$.
\end{theorem}

\begin{theorem}
\label{theorem3}
Under the regularity conditions~\ref{c:one} $-$~\ref{c:four}, the \cola\ estimator $\hatincthetak$  and the oracle estimator $\oracletheta$ are asymptotically equivalent, in the sense that $\Vert  \hatincthetak-\oracletheta \Vert^2=o_p(N_k^{-1})$ as $N_k \to \infty$.
\end{theorem}

Theorem \ref{theorem3} implies that the asymptotic difference between  $\hatincthetak$ and $\oracletheta$ is  $o_p({N_k}^{-1/2})$, and thus they are stochastically equivalent in the sense that they have the same asymptotic normal distribution. This implies that the \cola\ estimator is fully efficient.

\section{Simulation Experiments}\label{sec:simulation}
We evaluate the finite-sample performance of the proposed collaborative causal inference method, comparing the above \threeR, \twoR, and \oneR\ procedures with the classical meta-analysis (the inverse-variance weighted meta method \citep{cochran1954combination}) and the oracle estimation (i.e.,  the gold standard obtained by the centralized analysis). To mimic the motivating data example of the ITP-NODAT trial, we consider a binary outcome and estimate the causal log odds ratio. Additional simulations for continuous and count outcomes are included in the supplementary material. We first generate the full data 
under an assumed model and then split the data into $5$ subsets, one for a study site. We include three continuous variables $X_1, X_2 $, and $X_3$ independently drawn from  standard normal distribution $N(0,1)$, and two independent binary covariates $X_4$ and $X_5$ from  Bernoulli distribution with success probability of 0.5 and 0.6, respectively. We use $X$ to denote the vector of all five covariates. The treatment, $A$, follows  Bernoulli distribution with  $\text{pr}(A=1\mid X)=\mbox{expit}(0.5+0.3 X_1+ 0.3X_2+0.5X_3+0.5X_4+0.3X_5)$, where $\mbox{expit}(x) = 1/(1+e^{-x}) $.
 The outcome, $Y$, is drawn from a Bernoulli distribution with $\text{pr}(Y=1\mid A,X)=\mbox{expit}(-2.75+0.4A+0.3X_1+ 0.5X_2+0.3X_3+0.3X_4 +0.5X_5)$. The causal log odds ratio is estimated as $0.364$ using the Monte-Carlo simulation of $1,000,000$ random samples, and the proportion of cases $(Y=1)$ is approximately $30\%$. To simulate the scenarios of both unequal and equal proportions of cases across sites, we generate  group indicators $ {I}(
j = 5)$ which follows the Bernoulli distribution with $\text{pr}\{ {I}(j=5)\mid Y\}=\mbox{expit}(a+bY)$, the probability that a sample belongs to the $5$th site. The parameters $a$ and $b$ are predetermined such that we control the $5$th site to have approximately  $50$ samples, of which $5\%$ or $30\%$ are cases. The sample size $n_5$ at the $5$th site may not be exactly $50$ because it is round to the next integer. We split the rest of the samples into sites $1$ to $4$ with sizes of $100, 80, 80$, and $100-n_5$ , respectively. We consider the following scenarios to illustrate the finite performances of the proposed procedures.

\begin{scenario}\label{scen:1}
Cases are unequally distributed among all sites, and the $5$th site has a small proportion of cases, namely $5$\%, while holding the overall cases rate constant.
\end{scenario}
\begin{scenario}\label{scen:2}
Cases are equally distributed among all sites at approximately $30$\% at each site.
\end{scenario}
\begin{scenario}\label{scen:3}
The number of sites $K$ varies to be $5, 10, \mbox{and }{15}$ and $n_5 = 50$ to mimic the setting in the motivating example.
\end{scenario}
Due to space limit, we report simulation results from a typical case under Scenarios~\ref{scen:1} and~\ref{scen:2} with sample sizes of $100, 80, 80, 50, 50$ at each of the five sites in Table \ref{tab:simulation}. Results from scenario~\ref{scen:3} are presented in the supplementary material. Simulation results are summarized over $30,000$ replications limited to those where all algorithms converge successfully in order to make a fair comparison to the classical meta and oracle estimations.
\begin{table}[]

\begin{center}
	\caption{\label{tab:simulation}
	Simulation results for both equal outcome prevalence across sites and unequal outcome prevalence across sites. 
	}{
\begin{tabular}{lccccc}
                                                                   & \multicolumn{5}{c}{$n=(100, 80, 80, 50, 50$),  $P(Y=1)  \approx 0.05$ at the $5$th site}                                                                         \\
\multicolumn{1}{c}{\textit{Methods}}                               & \textit{$\Fails(\%)$} & \textit{$\CP(\%)$}   & \textit{$\Abias \times 10^{-3}$} & \textit{$\ASE \times 10^{-3}$} & \textit{$\ESE \times10^{-3}$} \\
\cellcolor[HTML]{FFFFFF}{\color[HTML]{222222} \textit{Oracle}}     & $0.00                  $& $94.5$                 & $214$                       & $262$                            & $270  $                         \\
\cellcolor[HTML]{FFFFFF}{\color[HTML]{222222} \textit{$\threeR$}}  & $0.00$                  & $94.5$                 & $213$                       & $261 $                           & $268 $                          \\
\cellcolor[HTML]{FFFFFF}{\color[HTML]{222222} \textit{$\twoR$}}    & $0.01$                  & $93.0$                 & $213$                       & $244$                            & $268$                           \\
\cellcolor[HTML]{FFFFFF}{\color[HTML]{222222} \textit{$\twoRinf$}} & $0.00$                  & $93.2$                 & $222$                       & $260$                            & $282$                           \\
\cellcolor[HTML]{FFFFFF}{\color[HTML]{222222} \textit{$\oneR$}}    & $0.01$                  & $92.4$                 & $222$                       & $249$                            & $282$                           \\
\textit{Meta}                                                      & $58.49$                 & $91.6$                 & $237$                       & $261$                            & $297$                           \\
                                                                   & \multicolumn{1}{l}{}  & \multicolumn{1}{l}{} & \multicolumn{1}{l}{}      & \multicolumn{1}{l}{}           & \multicolumn{1}{l}{}          \\
                                                                   & \multicolumn{5}{c}{$n=(100, 80, 80, 50, 50) $ $P(Y=1) \approx 0.3$ at the $5$th site}                                                                          \\
\multicolumn{1}{c}{\textit{Methods}}                               & \textit{$\Fails(\%)$} & \textit{$\CP(\%)$}   & \textit{$\Abias \times 10^{-3}$} & \textit{$\ASE \times 10^{-3}$} & \textit{$\ESE \times10^{-3}$} \\
\cellcolor[HTML]{FFFFFF}{\color[HTML]{222222} \textit{Oracle}}     & $0.00$                  & $94.7$                 & $217$                       & $264$                            & $273$                           \\
\cellcolor[HTML]{FFFFFF}{\color[HTML]{222222} \textit{$\threeR$}}  & $0.00$                  & $94.7$                 & $216$                       & $262$                            & $271 $                          \\
\cellcolor[HTML]{FFFFFF}{\color[HTML]{222222} \textit{$\twoR$}}    & $0.05$                  & $94.8$                 & $216$                       & $262$                            & $271$                           \\
\cellcolor[HTML]{FFFFFF}{\color[HTML]{222222} \textit{$\twoRinf$}} & $0.00$                  & $93.7$                 & $225$                       & $262$                            & $283$                           \\
\cellcolor[HTML]{FFFFFF}{\color[HTML]{222222} \textit{$\oneR$}}    & $0.05$                  & $94.1$                 & $225$                       & $266$                            & $283$                           \\
\textit{Meta}                                                      & $3.86$                  & $91.3$                 & $238 $                      & $258$                            & $298$                          
\end{tabular}}

\end{center}
\begin{tablenotes}{\item \Fails\, the number of non-convergence for incremental methods and the traditional meta-analysis method over $30,000$ replications; \CP,  coverage probability; \Abias, average absolute bias; \ASE, median estimated standard error of the estimates; \ESE\, empirical standard error. 
}\end{tablenotes}

\end{table} 
We evaluate the estimation and inference performances for different methods of estimating the causal effects for Scenario~\ref{scen:1} in the first sub-table of Table~\ref{tab:simulation} and we make the following observations: (i) The classical meta-analysis suffers substantial numerical failures due to low proportions of cases at site $5$, evident by the fact that $58.49\%$ of the simulation replicates fail to reach convergence. 
In contrast, the rate of failures ($<0.01 \%$) of the proposed incremental methods is significantly less than the meta-analysis method. 
(ii) The oracle estimation and the \threeR\ estimation produce very close coverage probability (\CP) and average absolute bias (\Abias). 
 This fully confirms the theoretical results given in Section~\ref{sec:Asymptotics}. By contrast, the \CP\ of the meta-analysis estimation is at $91.6\% $, much lower than the $95$\% nominal level, and the \Abias\ and empirical standard error (\ESE) are  larger than those of the other competing methods. (iii) The average bias of \twoR\ is the same as that of \threeR\ due to the same fully updated estimate $\incgammaallthree$, , equivalently $\incgammaalltwo$,  being used to update the casual effect in  ``Round 2" of each method.
 
 In Scenario~\ref{scen:2}, in which site 5 has a comparable case rate to other sites ($30\%$), in addition to the observations we made above, the coverage probability of both \twoR\ and \oneR\ is close to that of \threeR\ and the oracle estimator. Overall, in both Scenarios~\ref{scen:1} and~\ref{scen:2}, little statistical efficiency is lost by \threeR\ in the estimation of causal log odds ratio compared to the oracle method even when the outcome is severely unevenly distributed. 

 The previous simulation results are conducted under a relatively ideal situation when we know which site deviates the most from the overall distribution and place the site with the worst data quality as the last site in the sequence of updating. To gain some insight into how the ordering of sites affects the results, we investigate a less desirable situation where we do not know \textit{a priori} which site has the highest likelihood of failure in convergence due to either rare outcomes or skewed covariates. We consider three additional scenarios.
 \begin{scenario}\label{scen:4}
 Vary $\text{pr}(X_2=1)$ at site $1$ to be rare such that site $1$ has a rare binary covariate.
\end{scenario}
 \begin{scenario}\label{scen:5}
Rearrange the order of operation in Scenario 4.
\end{scenario}
\begin{scenario}\label{scen:6}
Vary $\text{pr}(Y=1 )$ at site $1$ to be rare such that the first site has a rare outcome.

\end{scenario}

\begin{figure}
\includegraphics[width=16cm]{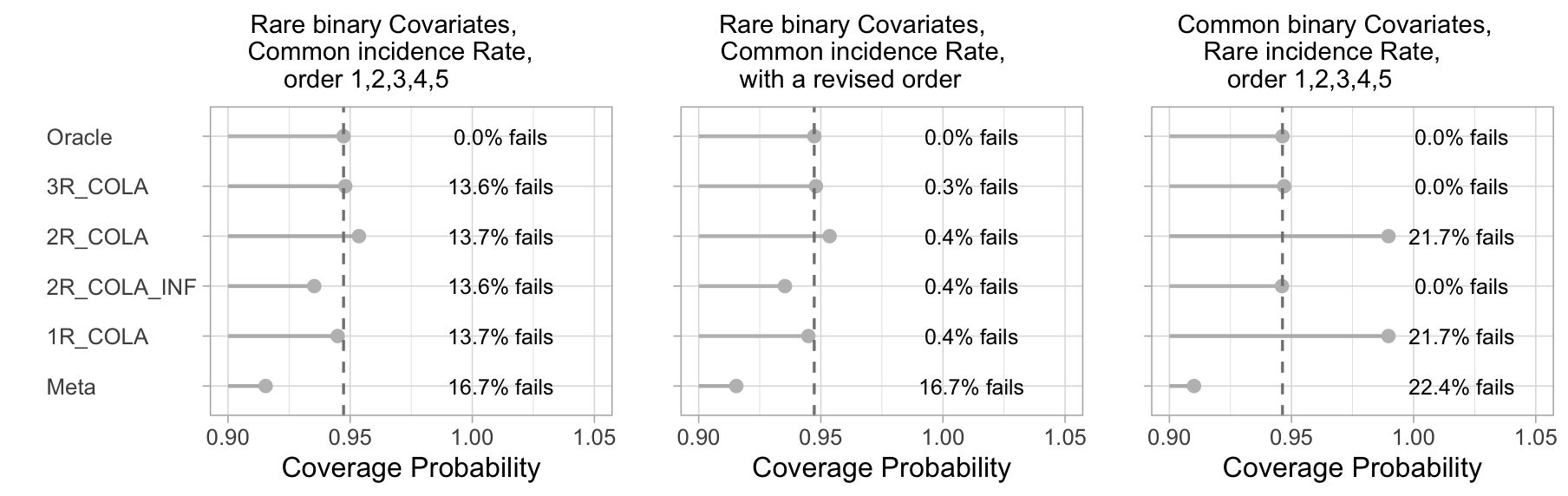}
	\caption{	\label{fig:rarecp}A comparison of coverage probabilities for all methods when the binary covariates or outcome incidence are rare based on 30000 replications. }

\end{figure}
When the binary $X_2$ is skewed at the first site as considered in Scenario~\ref{scen:4}, the local propensity score model may run into positivity assumption violations (Assumption~\ref{a1:pos}), and consequently, the $\PS$ model fitted within a local site may fail to converge or fail to generate stable and trustworthy results. The left panel in Fig~\ref{fig:rarecp} shows that $\cola$ and meta fail to converge $13\cdot6\%$ and $16\cdot7\%$ of the time respectively when the starting site has poor quality data. To address this issue, one may rearrange the order of sites. In Scenario~\ref{scen:5}, we switch bad starting sites that violate the positivity assumption with the latter good large sites that satisfy this assumption, so that the first ``good" site after excluding the one that violates the positivity assumption is the new starting site. All \cola\ methods including $\oneR$, $\twoR$, and $\threeR$ perform well and reach $\CP$ close to the oracle after the rearrangement due to fully updated propensity score model estimates are used for causal effect estimation and inference. This reordering strategy is not applicable to the meta-analysis method because it does not yield viable local results. 
 
In Scenario~\ref{scen:6} in which the positive case rate is low at the first site, a local treatment effect estimate is likely to be unstable. It is interesting to note that, once the \cola\ updating procedure continues and incorporates data from more sites, the incrementally updated point estimate becomes close to the true value. $\threeR$ performs equally well in comparison to the oracle method as shown in the right panel of Fig \ref{fig:rarecp}. 
 
The simulation results above lead to the practical guidelines summarized as follows. In general, when using $\cola$,  we suggest starting with the largest site to take advantage of large sample properties. If the outcome is rare or binary covariates are unbalanced at the largest site, then one can choose \threeR, or combine smaller sites with less severe distribution imbalances to create a new starting site and then proceed with either \twoR\  or \oneR. \threeR\ is the top choice if the communication cost is allowed as it provides higher statistical accuracy.

\section{Data Application Example}\label{sec:data application}
We apply the proposed collaborative inference method to analyze data from the Insulin Therapy for the Prevention of New-Onset Diabetes after Transplantation (ITP-NODAT) trial~\citep{Schwaiger_2021}. Two hundred and thirty-six kidney-transplantation patients are recruited in the study and randomized within each hospital to receive a diabetics preventive treatment, namely basal insulin injection right after kidney transplantation.  The primary goal of the ITP-NODAT trial is to estimate the effectiveness of basal insulin intervention in preventing overt post-transplantation diabetes mellitus (PTDM) at month $12$ after the randomization. Following the original analysis conducted by~\cite{Schwaiger_2021}, we will estimate the intent-to-treat effect of the basal insulin treatment. The case is defined as $Y=1$ if a patient receives antidiabetic therapy (which indicates the occurrence of diabetes), has $2$-hour plasma glucose $ \geq200$ mg/dL, or has Hemoglobin A1C (HbA1c) $\geq6.5$\%; otherwise $Y=0$.  
We adjust for the following covariates: age, gender, family history of diabetes, whether the living donor or deceased donor, whether first-time transplantation or repeated transplantation, whether having polycystic kidney diseases, and whether having glomerular diseases. The proportion of missingness in the covariates is mild, thus we conduct a complete-case analysis excluding $42$ dropouts and $8$ participants following \cite{Schwaiger_2021}.

We conduct \cola\ implementing the \twoR\ algorithm without requiring subject-level data sharing. This analysis is particularly meaningful for the ITP-NODAT trial because pooling data from the four hospitals took three years to complete due to various cross-country data-sharing barriers.  We also perform a centralized analysis of the pooled data as the gold standard for benchmarking, as well as the classical meta-analysis based on site-specific estimates for comparison.  

 The biggest challenge in the data analysis pertains to the unequal proportions of PTDM cases across hospitals, as shown in Section $4$ of the supplementary material. There were zero PTDM cases recorded in the treatment group at the Barcelona hospital and zero PTDM cases in the control group at the Graz hospital, which prevent us from getting any site-specific results at Barcelona and Graz, leading to an exclusion of two out of four site-specific estimates. This data attrition is undesirable in classical meta-analysis.

\begin{table}[]

\caption{Propensity score estimates for basal insulin treatment from combined data via centralized analysis and collaborative inference method. }{
\begin{tabular}{lrllrll}& \multicolumn{6}{c}{}                                                                          \\  &\multicolumn{3}{l}{``Gold-standard'' from combined data}                         & \multicolumn{3}{l}{Collaborative inference method}                       \\
\multicolumn{1}{c}{}                                                 & Estimates & \multicolumn{1}{c}{Std.Errors} & \multicolumn{1}{l}{p-value} & Estimates & \multicolumn{1}{c}{Std.Errors} & \multicolumn{1}{c}{p-value} \\
Gender                                                               & $0.13$      & $0.30$                           & $0.72$                        & $0.13$      & $0.30$                           & $0.67$                        \\
Age                                                                  & $-0.01$     & $0.01 $                          & $0.39 $                       & $-0.01$     & $0.01 $                          & $0.46 $                       \\
\begin{tabular}[c]{@{}l@{}}Family diabetes\\ history\end{tabular}    & $0.72$      & $0.44$                           & $0.10$                        & $0.72$      & $0.44$                           & $0.10$                        \\
First transplant                                                     & $0.03$      & $0.46$                           & $0.97$                        & $0.02$      &$0.46$                           & $0.95 $                       \\
\begin{tabular}[c]{@{}l@{}}Glomerular \\ disease\end{tabular}        & $-0.33$     & $0.32$                           & $0.31$                        & $-0.33$     & $0.32$                           & $0.32$                        \\
\begin{tabular}[c]{@{}l@{}}Polycystic kidney \\ disease\end{tabular} & $-0.41$     & $0.36$                           & $0.24$                        & $-0.42$     & $0.35$                           & $0.25$                        \\
Living Doner                                                         & $0.86$      & $0.45$                           & $0.05$                        & $0.87$      & $0.45$                           & $0.05$                        \\ 
\end{tabular}}
\label{tab:ps-table}
\begin{tablenotes}
\item We use \twoR\ algorithm for estimating collaborative inference ``global" PS parameters. 
\end{tablenotes}

\end{table}

Table \ref{tab:ps-table} presents the estimated coefficients in the propensity score model obtained from the \cola\ method based on summary statistics and the centralized method based on the pooled data. 
Then we obtain an inverse probability of treatment weighted estimate of the causal odds ratio and its $95\%$ confidence interval (CI). 
The estimated causal odds ratio  is $0.37\ (\text{96\% CI: }0.15, 0.93)$, which is very similar to the gold standard, $0.37 \ (\text{96\% CI: } 0.15, 0.91)$. In contrast, the classical meta-analysis is based on site-specific estimates from two out of four study sites due to rare outcomes, 
and the meta-estimated odds ratio is $0.62\ (\text{96\% CI: }0.20, 1.93)$ which overestimates the treatment effect by twice than the one from the centralized analysis. 

It is evident from our analysis that the basal insulin treatment reduces occurrences of PTDM with an estimated odds ratio that is significantly less than one. This is in agreement with the findings reported in \cite{Schwaiger_2021}, which performed a standard clinical trial analysis with no considerations of propensity score weighting. Little loss of statistical power occurred in our collaborative inference approach, while thoroughly overcoming data sharing barriers and enjoying the maximal protection of data privacy. Had our method and analysis been available, these important clinical findings could be published a few years earlier to add a new clinical treatment that benefits transplant patients. It is also worth noting that our proposed causal inference approach is not affected by imbalanced distributions of disease cases across study sites, a striking advantage over the classical meta-analysis method.

\section{Discussion} \label{sec: discussion}
In this paper, we introduce a collaborative operation of linked analysis (\cola) framework that overcomes data-sharing barriers and provides an efficient population average treatment effect estimate. We show the desirable asymptotic properties of the proposed distributed inference method. We also investigate the finite-sample performance through numerical experiments with four algorithms to implement \cola\ at different levels of communication costs. We find that little statistical efficiency is lost compared to the centralized method when estimating causal log odds ratios incrementally via \threeR\  and \twoR\ procedures. Even when the outcome is severely rare, \threeR\ achieves similar results as the oracle method. Although we focus our attention on binary outcomes, our \cola\ framework enjoys the same properties and performance in the numerical illustrations for other outcome types under the generalized linear models as shown in Section 4 of the supplementary material. 

Meta-analytic types of causal inference methods in certain parallel computing diagrams can fail at two levels: local sites fail to converge and thus the pooled inference results fail to reflect the true parameters of the underlying population. Convergence failures occur when some sites do not have enough variability in outcome measurements. In practice, our \cola\ methods only require the first site to have enough data variability which makes it an appealing method for multi-center clinical trials that involve small study sites. 
To facilitate \cola\ in practice, we provide an R package for data analysis and an interactive information communication platform that allows each site to run the R program independently and upload and download the summary statistics via our platform. Meanwhile, the emergence and development of federated learning which allows individual sites to collaboratively conduct data analysis while mitigating data privacy risks can be another possible solution to help us build an automated privacy-preserving software~\citep{li2020federated,fed_review,li2020review}. 

The current \cola\ framework is developed under a set of reasonable assumptions for identifiability and large-sample properties. Future work can potentially relax those assumptions and focus on examining the model's robustness in estimation by flagging out incompatible or outlying local data sites and reducing communication burdens between sites by allowing a varying control of data privacy in different parallel problems. In recent causal inference method development, a line of research is primarily aimed at combining multiple datasets collected by different designs from potentially heterogeneous populations~\citep{Yang_Ding_2020, Wang_Lu_Chen_Li_Tiwari_Xu_Yue_2020, Bareinboim_Pearl_2016,shi2021data}. Most data fusion methods estimate causal treatment effect by incorporating patient-level data from auxiliary data sources into the main data source without consideration of data privacy issues. We plan to take advantage of the privacy-preserving nature of \cola\ and extend it to data fusion problems. Another interesting potential extension of our method is to transport our \cola\ estimation which targets the population underlying the current multi-center clinical trials to a new  population~\citep{dahabreh2020toward,han2021federated}.

	\clearpage
	\spacingset{1.45}
	\bibliographystyle{agsm}
	\bibliography{SC}

@article{dahabreh2020toward,
  title={Toward causally interpretable meta-analysis: Transporting inferences from multiple randomized trials to a new target population},
  author={Dahabreh, Issa J and Petito, Lucia C and Robertson, Sarah E and Hern{\'a}n, Miguel A and Steingrimsson, Jon A},
  journal={Epidemiology},
  volume={31},
  number={3},
  pages={334--344},
  year={2020},
  publisher={Wolters Kluwer}
}

@article{hernan2002estimating,
  title={Estimating the causal effect of zidovudine on CD4 count with a marginal structural model for repeated measures},
  author={Hern{\'a}n, Miguel A and Brumback, Babette A and Robins, James M},
  journal={Statistics in medicine},
  volume={21},
  number={12},
  pages={1689--1709},
  year={2002},
  publisher={Wiley Online Library}
}

@article{Schwaiger_2021, title={Early Postoperative Basal Insulin Therapy versus Standard of Care for the Prevention of Diabetes Mellitus after Kidney Transplantation: A Multicenter Randomized Trial}, volume={32}, ISSN={1046-6673}, DOI={10.1681/ASN.2021010127}, number={8}, journal={Journal of the American Society of Nephrology}, publisher={American Society of Nephrology}, author={Schwaiger, Elisabeth and Krenn, Simon and Kurnikowski, Amelie and Bergfeld, Leon and Pérez-Sáez, María José and Frey, Alexander and Topitz, David and Bergmann, Michael and Hödlmoser, Sebastian and Bachmann, Friederike and et al.}, year={2021}, pages={2083–2098} }

@article{rosenbaum1983central,
  title={The central role of the propensity score in observational studies for causal effects},
  author={Rosenbaum, Paul R and Rubin, Donald B},
  journal={Biometrika},
  volume={70},
  number={1},
  pages={41--55},
  year={1983},
  publisher={Oxford University Press}
}

@article{Stefanski_Boos_2002, title={The Calculus of M-Estimation}, volume={56}, ISSN={0003-1305}, number={1}, journal={The American Statistician}, publisher={[American Statistical Association, Taylor & Francis, Ltd.]}, author={Stefanski, Leonard A. and Boos, Dennis D.}, year={2002}, pages={29–38} }

@article{neyman1923application,
  title={On the application of probability theory to agricultural experiments. essay on principles. section 9.(tlanslated and edited by dm dabrowska and tp speed, statistical science (1990), 5, 465-480)},
  author={Neyman, Jerzy S},
  journal={Annals of Agricultural Sciences},
  volume={10},
  pages={1--51},
  year={1923}
}

@article{cochran1954combination,
  title={The combination of estimates from different experiments},
  author={Cochran, William G},
  journal={Biometrics},
  volume={10},
  number={1},
  pages={101--129},
  year={1954},
  publisher={JSTOR}
}

@book{mccullagh2019generalized,
  title={Generalized linear models},
  author={McCullagh, Peter and Nelder, John A},
  year={2019},
  publisher={Routledge}
}

@article{Bareinboim_Pearl_2016, title={Causal inference and the data-fusion problem}, volume={113}, ISSN={0027-8424, 1091-6490}, DOI={10.1073/pnas.1510507113}, number={27}, journal={Proceedings of the National Academy of Sciences}, publisher={National Academy of Sciences}, author={Bareinboim, Elias and Pearl, Judea}, year={2016}, month={Jul}, pages={7345–7352} }

@article{Wang_Lu_Chen_Li_Tiwari_Xu_Yue_2020, title={Propensity score-integrated composite likelihood approach for incorporating real-world evidence in single-arm clinical studies}, volume={30}, ISSN={1054-3406}, DOI={10.1080/10543406.2019.1684309}, abstractNote={In medical product development, there has been an increased interest in utilizing real-world data which have become abundant with recent advances in biomedical science, information technology, and engineering. High-quality real-world data may be analyzed to generate real-world evidence that can be utilized in the regulatory and healthcare decision-making. In this paper, we consider the case in which a single-arm clinical study, viewed as the primary data source, is supplemented with patients from a real-world data source containing both clinical outcome and covariate data at the patient-level. Propensity score methodology is used to identify real-world data patients that are similar to those in the single-arm study in terms of the baseline characteristics, and to stratify these patients into strata based on the proximity of the propensity scores. In each stratum, a composite likelihood function of a parameter of interest is constructed by down-weighting the information from the real-world data source, and an estimate of the stratum-specific parameter is obtained by maximizing the composite likelihood function. These stratum-specific estimates are then combined to obtain an overall population-level estimate of the parameter of interest. The performance of the proposed approach is evaluated via a simulation study. A hypothetical example based on our experience is provided to illustrate the implementation of the proposed approach.}, number={3}, journal={Journal of Biopharmaceutical Statistics}, publisher={Taylor & Francis}, author={Wang, Chenguang and Lu, Nelson and Chen, Wei-Chen and Li, Heng and Tiwari, Ram and Xu, Yunling and Yue, Lilly Q.}, year={2020}, month={May}, pages={495–507} }

@article{Yang_Ding_2020, title={Combining Multiple Observational Data Sources to Estimate Causal Effects}, volume={115}, ISSN={0162-1459, 1537-274X}, DOI={10.1080/01621459.2019.1609973}, abstractNote={The era of big data has witnessed an increasing availability of multiple data sources for statistical analyses. We consider estimation of causal effects combining big main data with unmeasured confounders and smaller validation data with supplementary information on these confounders. Under the unconfoundedness assumption with completely observed confounders, the smaller validation data allow for constructing consistent estimators for causal effects, but the big main data can only give error-prone estimators in general. However, by leveraging the information in the big main data in a principled way, we can improve the estimation efficiencies yet preserve the consistencies of the initial estimators based solely on the validation data. Our framework applies to asymptotically normal estimators, including the commonly used regression imputation, weighting, and matching estimators, and does not require a correct specification of the model relating the unmeasured confounders to the observed variables. We also propose appropriate bootstrap procedures, which makes our method straightforward to implement using software routines for existing estimators. Supplementary materials for this article are available online.}, number={531}, journal={Journal of the American Statistical Association}, author={Yang, Shu and Ding, Peng}, year={2020}, month={Jul}, pages={1540–1554} }

@article{Zhou_2017, title={Sentinel Modular Program for Propensity Score-Matched Cohort Analyses: Application to Glyburide, Glipizide, and Serious Hypoglycemia}, volume={28}, ISSN={1531-5487}, DOI={10.1097/EDE.0000000000000709}, number={6}, journal={Epidemiology (Cambridge, Mass.)}, author={Zhou, Meijia and Wang, Shirley V. and Leonard, Charles E. and Gagne, Joshua J. and Fuller, Candace and Hampp, Christian and Archdeacon, Patrick and Toh, Sengwee and Iyer, Aarthi and Woodworth, Tiffany Siu and et al.}, year={2017}, month={Nov}, pages={838–846} }

@article{Luo_2020,
  title={Renewable estimation and incremental inference in generalized linear models with streaming data sets},
  author={Luo, Lan and Song, Peter X-K},
  journal={Journal of the Royal Statistical Society: Series B (Statistical Methodology)},
  volume={82},
  number={1},
  pages={69--97},
  year={2020},
  publisher={Wiley Online Library}
}

@article{carter2016,
  title={Avoiding pitfalls with implementation of randomized controlled multicenter trials: strategies to achieve milestones},
  author={Carter, Barry L and Ardery, Gail},
  journal={Journal of the American Heart Association},
  volume={5},
  number={12},
  pages={e004432},
  year={2016},
  publisher={Am Heart Assoc}
}

@article{coates2020,
  title={Challenges associated with managing a multicenter clinical trial in severe burns},
  author={Coates, Elsa C and Mann-Salinas, Elizabeth A and Caldwell, Nicole W and Chung, Kevin K},
  journal={Journal of Burn Care \& Research},
  volume={41},
  number={3},
  pages={681--689},
  year={2020},
  publisher={Oxford University Press US}
}

@misc{mello2013,
  title={Preparing for responsible sharing of clinical trial data},
  author={Mello, Michelle M and Francer, Jeffrey K and Wilenzick, Marc and Teden, Patricia and Bierer, Barbara E and Barnes, Mark},
  year={2013},
  publisher={Mass Medical Soc}
}

@article{rubin2005,
  title={Causal inference using potential outcomes: Design, modeling, decisions},
  author={Rubin, Donald B},
  journal={Journal of the American Statistical Association},
  volume={100},
  number={469},
  pages={322--331},
  year={2005},
  publisher={Taylor \& Francis}
}

@book{song2007,
  title={Correlated data analysis: modeling, analytics, and applications},
  author={Song, Peter X-K},
  year={2007},
  publisher={Springer Science \& Business Media}
}

@article{li2020federated,
  title={Federated learning: Challenges, methods, and future directions},
  author={Li, Tian and Sahu, Anit Kumar and Talwalkar, Ameet and Smith, Virginia},
  journal={IEEE Signal Processing Magazine},
  volume={37},
  number={3},
  pages={50--60},
  year={2020},
  publisher={IEEE}
}

@article{fed2021,
  author    = {Ruoxuan Xiong and
               Allison Koenecke and
               Michael Powell and
               Zhu Shen and
               Joshua T. Vogelstein and
               Susan Athey},
  title     = {Federated Causal Inference in Heterogeneous Observational Data},
  journal   = {CoRR},
  volume    = {abs/2107.11732},
  year      = {2021},
  url       = {https://arxiv.org/abs/2107.11732},
  eprinttype = {arXiv},
  eprint    = {2107.11732},
  bibsource = {dblp computer science bibliography, https://dblp.org}
}

@article{tsiatis2006,
  title={Semiparametric theory and missing data},
  author={Tsiatis, Anastasios A},
  year={2006},
  publisher={Springer}
}

@article{duan2019heterogeneity,
  title={Heterogeneity-aware and communication-efficient distributed statistical inference},
  author={Duan, Rui and Ning, Yang and Chen, Yong},
  journal={arXiv preprint arXiv:1912.09623},
  year={2019}
}

@inproceedings{duan2018odal,
  title={ODAL: A one-shot distributed algorithm to perform logistic regressions on electronic health records data from multiple clinical sites},
  author={Duan, Rui and Boland, Mary Regina and Moore, Jason H and Chen, Yong},
  booktitle={BIOCOMPUTING 2019: Proceedings of the Pacific Symposium},
  pages={30--41},
  year={2018},
  organization={World Scientific}
}

@article{jordan2018communication,
  title={Communication-efficient distributed statistical inference},
  author={Jordan, Michael I and Lee, Jason D and Yang, Yun},
  journal={Journal of the American Statistical Association},
  year={2018},
  publisher={Taylor \& Francis}
}

@article{robins1994estimation,
  title={Estimation of regression coefficients when some regressors are not always observed},
  author={Robins, James M and Rotnitzky, Andrea and Zhao, Lue Ping},
  journal={Journal of the American statistical Association},
  volume={89},
  number={427},
  pages={846--866},
  year={1994},
  publisher={Taylor \& Francis}
}

@article{godambe1960,
  title={An optimum property of regular maximum likelihood estimation},
  author={Godambe, Vidyadhar P},
  journal={The Annals of Mathematical Statistics},
  volume={31},
  number={4},
  pages={1208--1211},
  year={1960},
  publisher={JSTOR}
}

@book{godambe1991,
  title={Estimating functions},
  author={Godambe, Vidyadhar P},
  year={1991},
  publisher={Oxford University Press}
}

@article{fed_review,
  title={Advances and open problems in federated learning},
  author={Kairouz, Peter and McMahan, H Brendan and Avent, Brendan and Bellet, Aur{\'e}lien and Bennis, Mehdi and Bhagoji, Arjun Nitin and Bonawitz, Kallista and Charles, Zachary and Cormode, Graham and Cummings, Rachel and others},
  journal={Foundations and Trends{\textregistered} in Machine Learning},
  volume={14},
  number={1--2},
  pages={1--210},
  year={2021},
  publisher={Now Publishers, Inc.}
}

@article{li2020review,
  title={A review of applications in federated learning},
  author={Li, Li and Fan, Yuxi and Tse, Mike and Lin, Kuo-Yi},
  journal={Computers \& Industrial Engineering},
  pages={106854},
  year={2020},
  publisher={Elsevier}
}

@article{shi2021data,
  title={Data Integration in Causal Inference},
  author={Shi, Xu and Pan, Ziyang and Miao, Wang},
  journal={arXiv preprint arXiv:2110.01106},
  year={2021}
}

@article{han2021federated,
  title={Federated Adaptive Causal Estimation (FACE) of Target Treatment Effects},
  author={Han, Larry and Hou, Jue and Cho, Kelly and Duan, Rui and Cai, Tianxi},
  journal={arXiv preprint arXiv:2112.09313},
  year={2021}
}

\end{document}